\documentstyle[twoside,fleqn,espcrc2]{article}

\input epsf.tex

\title{Experimental search for solar axions.}

\author{A.O. Gattone$^a$, D. Abriola$^a$, F.T. Avignone$^b$, R.L.
Brodzinski$^c$, J.I. Collar$^d$, R.\ J.\ Creswick$^b$, \\ D.E. Di
Gregorio$^a$, H.A. Farach$^b$, C.\ K.\ Gu\'erard$^{a,b}$,
 F. Hasenbalg$^a$, H. Huck$^a$, \\ H.S. Miley$^c$, A. Morales$^e$, J.
Morales$^e$, S.\ Nussinov$^f$, A. Ortiz de Sol\'orzano$^e$, \\ J.H.
Reeves$^c$, J.A. Villar$^e$, and K.\ Zioutas$^g$ (The SOLAX Collaboration)\\
\ \ \ \\
$^a$ Department of Physics, TANDAR Laboratory, CNEA, Buenos Aires, 
Argentina\thanks{Supported partially by grants from CONICET, Fundaci\'on ANTORCHAS and UNSAM} \\
\ \ \ \\
$^b$ Department of Physics, University of South Carolina,
Columbia, SC 29208, USA \\
\ \ \ \\
$^c$ Pacific Northwest National Laboratory, Richland, WA 99352, USA \\
\ \ \ \\
$^d$ CERN, CH-1211 Geneva, 23 Switzerland \\
\ \ \ \\
$^e$ Laboratorio de F\'{\i}sica Nuclear y Altas Energ\'{\i}as, 
Universidad de Zaragoza, Zaragoza, Spain \\
\ \ \ \\
$^f$ Department of Physics, Tel Aviv University, Tel Aviv, Israel \\
\ \ \ \\
$^g$ Department of Physics, University of Thessaloniki, GR54006 Thessaloniki, 
Greece}

\begin{document}

%\date{}
%\maketitle

\begin{abstract}
A new technique has been used to search for solar axions using a single crystal germanium detector. It exploits the coherent conversion of axions into photons when their angle of incidence satisfies a Bragg condition with a crystalline plane. The analysis of approximately 1.94 kg.yr of data from the 1 kg DEMOS detector in Sierra Grande, Argentina, yields a new laboratory bound on axion-photon coupling of $g_{a\gamma\gamma}<2.7\times 10^{-9}$ GeV$^{-1}$, independent of axion mass up to $\sim$ 1 keV.
\end{abstract}

\maketitle

\section{INTRODUCTION}

The axion is the Nambu-Goldstone boson of the broken chiral global U(1) symmetry introduced by Peccei and Quinn twenty years ago~\cite{pq} to dynamically solve the, so called, ``strong CP problem" of QCD. This prompted many theoretical investigations and experimental searches.

\begin{figure}[htb]
%\vspace{7cm}
\epsfxsize=8.0truecm
\epsffile{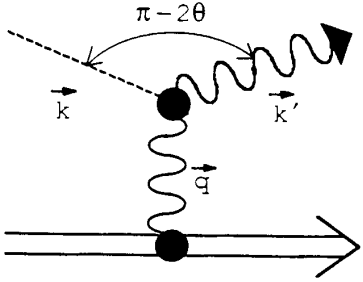}
\caption{\em Feynman diagram of the Primakoff conversion of axions into photons.}
\label{fig1}
\end{figure}

The axion-photon interaction Lagrangian is
${\cal{L}}_{int}=(1/4M)aF_{\mu\nu}F^{\alpha\beta}\varepsilon_{\mu\nu\alpha\beta}$
where $a$ is the pseudoscalar axion field, $F_{\mu\nu}$ is the
electromagnetic field, and $1/M = g_{a\gamma\gamma}$ is the axion-photon
coupling.  The objective of this experiment is to detect solar axions
through their coherent Primakoff conversion (see Fig.~\ref{fig1}) into
photons in the lattice of a germanium crystal when the incident angle
satisfies the Bragg condition.  As it turns out~\cite{cr}, the detection
rates in various energy windows are correlated with the relative
orientations of the detector and the sun.  This correlation results in
a temporal structure which should be a distinctive, unique signature of
the axion. We present here the
results of a search using a 1~kg, ultra--low background germanium
detector installed in the HIPARSA iron mine in Sierra Grande, Argentina
at 41$^{\rm o}$~41'~24''~S and 65$^{\rm o}$~22'~W.  A
complete description of the experimental set--up was given earlier by
Di Gregorio, et al.\cite{dg} and Abriola, et al. \cite{ab}.  This
experiment was motivated by earlier papers by Buchm\"uller and
Hoogeveen \cite{bh} and by Paschos and  Zioutas \cite{pz}; the
present technique was originally suggested by Zioutas and developed by
Creswick, et al. \cite{cr}.

The  terrestrial flux of axions from the sun can be approximated by the
expression \cite{bibber}:
\begin{equation}
  \frac{d\Phi}{dE}\;=\; \lambda^{1/2} \,
  \frac{\Phi_{\rm_o}(E/E_{\rm o})^3}{E_{\rm_o}(e^{E/E_{\rm o}}\,-\,1)} \; ,
\end{equation}
where $\lambda = (g_{a\gamma \gamma} \times 10^8)^4$ and is
dimensionless, $E_{\rm o} = 1.103$~keV, and $\Phi_{\rm o} = 5.95 \times
10^{14}$~cm$^{-2}$~sec$^{-1}$. The total flux for $\lambda = 1$
integrated from 0 to 12~keV is $3.54 \times
10^{15}$~cm$^{-2}$~sec$^{-1}$. The spectrum is a continuum peaking at
about 4~keV decreasing to a negligible contribution above 8~keV. The
differential
cross section for Primakoff conversion on an atom with nuclear charge
$Ze$ is:

\begin{equation}
    \frac{d\sigma}{d\Omega} \;=\; \left [
    \frac{Z^2 \alpha \hbar^2 c^2 g_{a\gamma \gamma}^2}{16 \pi} \right] \,
    \frac{q^2\,(4k^2 \,-\, q^2)}{(q^2\,+\,r_{\rm o}^{-2})^2} \; ,
\end{equation}
where $q$ is the momentum transfer, $k$ is the momentum of the incoming
axion, and $r_{\rm o}$ is the screening length of the atom
in the lattice.  For germanium, $\sigma_{\rm o} = Z^2 \alpha \hbar^2 c^2
g_{a\gamma \gamma}/8 \pi = 1.15 \times 10^{-44}$~cm$^2$ when
$g_{a\gamma \gamma} = 10^{-8}$~GeV$^{-1}$, or equivalently $\lambda =
1$.

For light axions the Primakoff process in a periodic lattice is
coherent when the Bragg condition ($2 d \sin\theta = n\lambda$)
is satisfied, that is when $\vec{q}$ transferred to the crystal is a
reciprocal lattice vector $\vec{G} = 2\pi(h,k,l)/a_{\rm o}$. Here
$a_{\rm o}$ is the size of the conventional cubic cell, and $h$, $k$,
and $l$ are integers.

It was shown that the rate of conversion of axions with energy $E$ when
the sun is in the direction $\hat{k}$, $\dot{N}(\hat{k},E)$, can be
written \cite{cr}:

\begin{eqnarray}
  \dot{N}(\hat{k},E) & = &  2\hbar c \,\frac{V}{v_c}\, \sum_G \left| S(G)
  \right|^2 \,
   \frac{d\sigma}{d\Omega}(\vec{G})\, \frac{1}{|\vec{G}|^2}\,
     \times \nonumber \\
     & & \hspace{1.cm} \frac{d\Phi}{dE} \delta(E-\frac{\hbar c |\vec{G}|^2}
     {2\hat{k}\cdot \vec{G}}) \; ,  \label{3}
\end{eqnarray}
where $V$ is the volume of the crystal, $v_c$ is the volume of a unit
cell, $S(G)$ is the structure function for germanium, and $d\Phi/dE$ is
evaluated at the axion energy of $\hbar c |\vec{G}|^2/ 2\hat{k} \cdot
\vec{G}$. The structure function for germanium is:
\begin{eqnarray}
  S(G) & = &\left[ 1 \,+\, e^{i\pi(h+k+l)/2} \right] \left[ 1
\,+\,e^{i\pi(h+k)} \,+\,  \right. \nonumber \\
  & &   \left. e^{i\pi(h+l)} \,+\, e^{i\pi(k+l)}
  \right] \; .
\end{eqnarray}
Note that in (\ref{3}) the coherent conversion of axions occurs only for a
particular axion energy given the position of the sun, $\hat{k}$, and
reciprocal lattice vector $\vec{G}$. However, the detector has a finite
energy resolution; for the detector in Sierra Grande it is 1~keV FWHM
at 10~keV. We take this into account by smoothing $\dot{N}(\hat{k},E)$
with a Gaussian of the appropriate width. Finally, we take the relevant
part of the energy spectrum, in this case from the threshold energy of
4~keV up to 8~keV (which is just below the X-rays at 10~keV), and
calculate the total rate of conversion in windows of width $\Delta E$,
typically 0.5~keV,
\begin{eqnarray}
 R(\hat{k},E) & = & 2\hbar c\,\frac{V}{v_c}\, \sum_G \left| S(G) \right|^2 
   \frac{d\sigma}{d\Omega} \, \frac{1}{ \left| \vec{G}\right|^2} \,
   \times \hspace{0.5cm} \nonumber \\
 & & \hspace{-1.9cm} \frac{d\Phi}{dE} \, \frac{1}{2} \left[{\rm erf} \left(\frac{E-E_a}
   {\sqrt{2}\sigma}\right) - {\rm erf}
   \left(\frac{E-E_a-\Delta E}
   {\sqrt{2}\sigma}\right)\right], \nonumber \\ & & \phantom{\left| S(G) \right|^2} \label{5}
\end{eqnarray}
where $ E_a=\hbar c \left| G \right|^2/(2\hat{k}\cdot \vec{G})$ and ${\rm 
 erf}(x)=\frac{2}{\sqrt{\pi}}\int_0^x e^{-t^2}dt$
is the error function. In equation
(\ref{5}) we have neglected the angular size of the core of the sun and
the mass of the axion which is justified when
$m_a c^2$ is small compared to the core temperature of the sun
\cite{bh}, i.e., up to a few keV.

The theoretical axion detection rate for this detector, calculated with
equation (\ref{5}), is shown in Figure \ref{fig2}.  The position of the
sun is computed at any instant in time using the U.S. Naval Observatory
Subroutines (NOVAS) \cite{novas}.  The pronounced variation in
$R(\hat{k},E)$ as a function of time invites the data to be analyzed
with the correlation function:

\begin{figure}[htb]
%\vspace{7cm}
\epsfxsize=8.0truecm
\epsffile{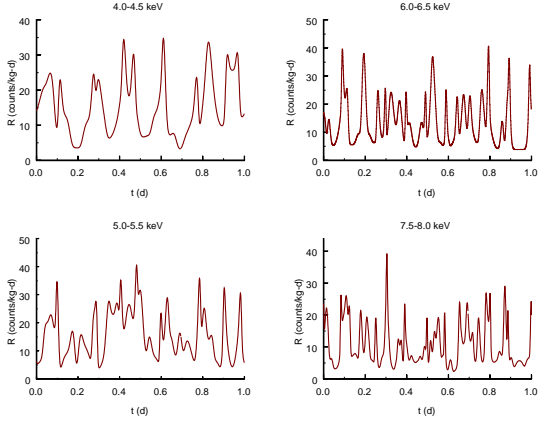}
\caption{\em A typical axion-photon conversion rate, $R(t,E)$, for various energy bands. The experimental energy resolution FWHM=1.0 keV at 10 keV was used.}
\label{fig2}
\end{figure}
\begin{equation}
   \chi\; \equiv \; \sum_{i=1}^n\, \left[
       R(t_i,E)\; -\, \langle R(E)\rangle \right] \, n(t_i) \; ,
\end{equation}
where $R(t_i,E)$ is the smooth shape of the theoretical rate at the
instant of time, $t_i$, $\langle R(E)\rangle$ is the average rate over
a finite time interval, and $n(t_i)$ is the number of events at $t_i$
in a time interval $\Delta t$, usually 0 or 1. The choice for the
weighting function $W(t,E)=R[\hat{k}(t),E] - \langle
R(E)\rangle$ is motivated by the requirement that any constant
background average to zero in $\chi$, whereas a counting rate which
follows $R[\hat{k}(t),E]$ increases $\chi$.

The number of counts at time, $t$, in the interval $\Delta t$
is assumed to be due in part to axions and in part to background
governed by a Poisson process with mean:

\begin{equation}
\langle n(t) \rangle\; =\; \left[ \lambda\, R(t,E)\, +\,b(E) \right]\, 
  \Delta t, \label{7}
\end{equation}
where $b(E)$ is constant in time.

The average value of $\chi$ is then,

\begin{eqnarray}
\langle \chi\rangle & = & \sum_i\left[ R(t_i,E) \; -\, 
   \langle R(E) \rangle \right]\left[ \lambda R(t_i,E) \, + \, 
\right. \nonumber \\
& & \hspace{2.0cm} \left. b(E) \right]\Delta t  \nonumber \\
   & \; =\; & 
   \sum_i W(t_i,E) \left[ \lambda R(t_i,E) +b(E) \right]\, 
   \Delta t . \label{8}
\end{eqnarray}

We can add and subtract the constant quantity $\lambda \langle R(E)
\rangle$ to the second factor in eq.~(\ref{8}). Any time independent
contributions multiplied by $W(t,E)$ in eq.~(\ref{8}), and summed over time,
will vanish.  Accordingly, taking the limit as $\Delta t \rightarrow
0$, we obtain:

\begin{equation}
 \langle \chi(\lambda) \rangle\; =\; \lambda \, \int_0^T\, W^2(t,E)\, dt.
\end{equation}
The expected uncertainty in $\chi$, $(\Delta \chi)^2 = \langle \chi^2
\rangle - \langle \chi \rangle^2$, is given by,
\begin{eqnarray}
(\Delta \chi)^2 & = & \sum_i\, \sum_j \,
W(t_i,E) W(t_j,E)\,  \times \nonumber \\
  &   &  \left[\langle n(t_i) n(t_j) \rangle \, -\,
  \langle n(t_i) \rangle \langle n(t_j) \rangle \right]
   \nonumber \\
  & = & \sum_i  W^2(t_i,E) 
  \left[\langle n(t_i)^2 \rangle -\langle n(t_i) \rangle^2 \right], 
\nonumber \\
\end{eqnarray}
where the square bracket is $\langle \Delta n(t_i) \rangle^2$, which in
Poisson statistics is $\langle n(t_i)\rangle$. Accordingly,
\begin{equation}
   (\Delta \chi)^2 \;=\; \sum_{i}\, W^2(t_i,E) \langle n(t_i) \rangle.
\end{equation}
By (\ref{7}) we have:
\begin{eqnarray}
  (\Delta \chi)^2  &= & \sum_{i}\, W^2(t_i,E)\, \left[ \lambda\,
  R(t_i,E)\, +\,b(E) \right]\, \Delta t  \nonumber \\
 & = & \sum_{i}  W^2(t_i,E) 
 \left\{ \lambda \left[ R(t_i,E) -
  \langle R(E) \rangle \right] \right. \nonumber \\
 & & + \left. \lambda \langle R(E) \rangle  + 
   b(E) \right\} \Delta t,
\end{eqnarray}
which in the limit $\Delta t \rightarrow 0$ becomes,
\begin{eqnarray}
   (\Delta \chi)^2 & = &\lambda  \int_0^T W^3(t,E) dt \, + \, \nonumber \\
   & & R_T(E)  \int_0^T W^2(t,E) dt.
\end{eqnarray}
The quantity $ R_T(t,E)$ is the average total counting rate, including both
axion conversions and background.

The data are separately analyzed in energy bins, $\Delta E_k$, fixed by
the detector resolution (FWHM $\sim 1$~keV in this case). The
likelihood function is then constructed:

\begin{equation}
L(\lambda) \;=\; \prod_k\, \exp \left[
   \frac{-(\chi_k \,-\, \langle \chi_k \rangle )^2}
    {2 (\Delta \chi_k)^2} \right].
\end{equation}
To an excellent approximation $(\Delta \chi_k)^2$ is dominated by
background. Maximizing the likelihood function, the most probable
value of $\lambda$ is given by

\begin{equation}
  \lambda_{\rm o}\; =\; \sum_k \, \chi_k \,/\, \sum_k\, A_k,
\end{equation}
where,
\begin{equation}
  A_k\; \equiv\; \int_0^T\, W_k^2(t,E)\, dt ,
\end{equation}
and the width of the likelihood function is given by,
\begin{equation}
  \sigma_{\lambda} \;=\; \left( \sum_k \, A_k\,/\, b_k \right)^{-1/2}.
\end{equation}
We note that $A_k$ is proportional to the time of the experiment, so
that $\sigma_{\lambda}$ decreases as $T^{1/2}$. The background scales
with the detector mass, while $A_k$ scales as the square of the
detector mass, therefore $\sigma_{\lambda}$ decreases as $(M_d
T)^{-1/2}$.

We have carried out extensive Monte-Carlo test of this method of
analysis. As a test of our analysis, typical results for the likelihood
function for the cases $\lambda = 0$ (no axions) and $\lambda = 0.003$
were calculated with realistic backgrounds for a detector operating
with the same mass, energy resolution, and threshold as the DEMOS
detector at the latitude and longitude of Sierra Grande for one year.
It is clear from this calculation that the correlation function
analysis is consistent and quite sensitive to the presence of a
variation in the counting range due to solar axions with a signal to
noise ratio less than 1\%.

\begin{figure}[htb]
%\vspace{7cm}
\epsfxsize=8.0truecm
\epsffile{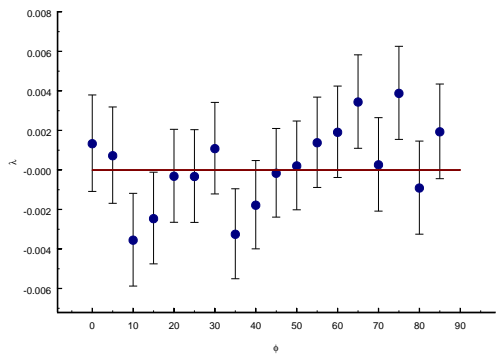}
\caption{\em Values of $\lambda$ calculated from the 707 days of data as a function of the azimuthal angle $\phi$. The error bars are $1 \sigma$.}
\label{fig3}
\end{figure}

The vertical axis of our detector is the (100) crystalline axis.  The
orientation of the (010) and (001) axes are unknown at this time.
Therefore, to place a bound on the axion interaction rate, the data
must be analyzed for many azimuthal orientations of the crystal, and
the weakest bound selected.  The results of these calculations for 707~days of
data in the energy range from 4 to 8~keV in 0.5~keV intervals are shown
in Figure \ref{fig3}.

The most conservative upper bound on $\lambda$, or equivalently $g_{a
\gamma \gamma}$, is found by taking for $\phi$ the angle at which
$\lambda$ is maximum. This yields an upper bound on the axion--photon
coupling constant $g_{a \gamma \gamma} < 2.7 \times 10^{-9}$~GeV$^{-1}$
at the 95\% confidence level.

\begin{figure}[htb]
%\vspace{7cm}
\epsfysize=7.0truecm
\epsfxsize=8.0truecm
\epsffile{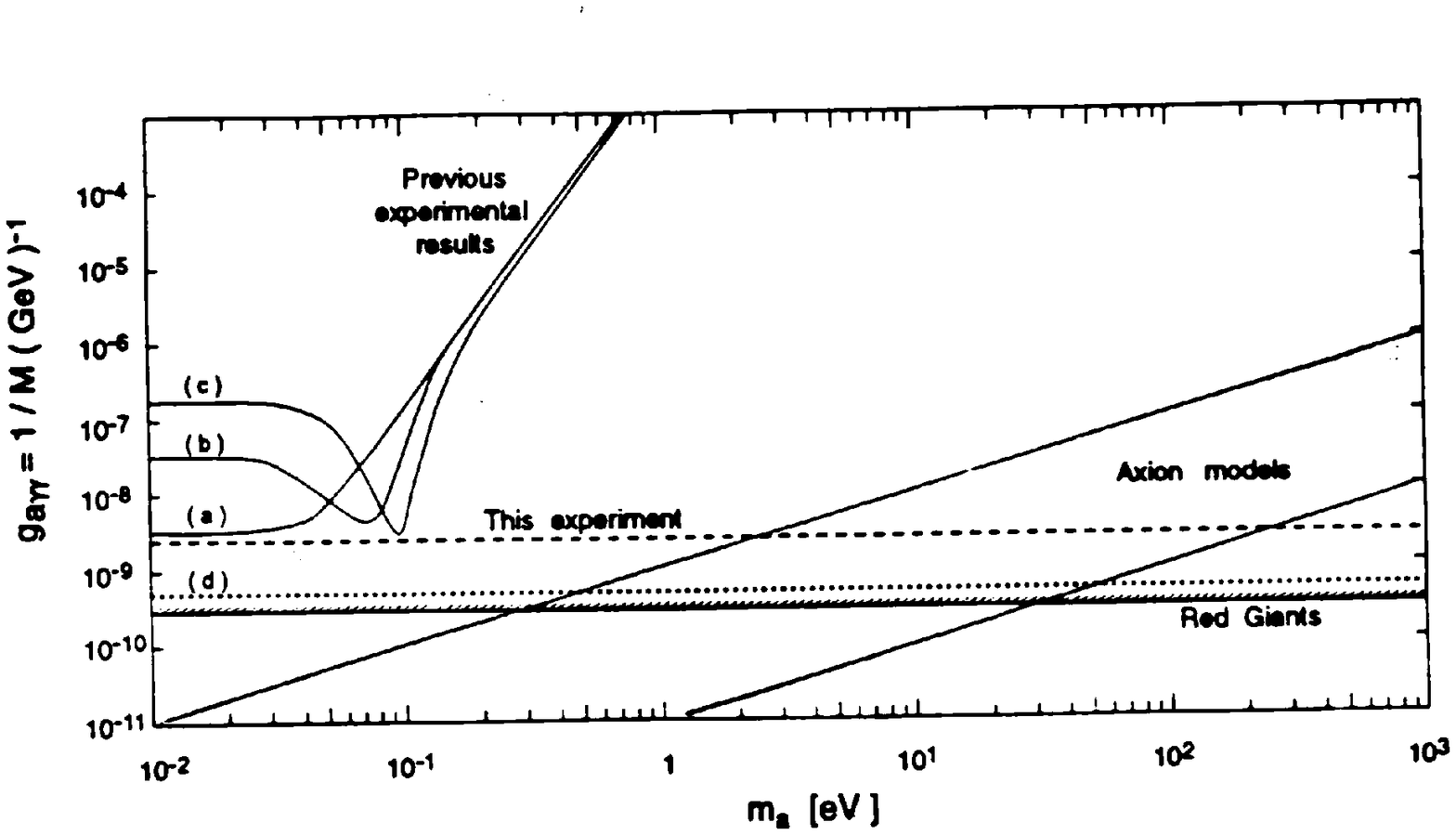}
\caption{\em Exclusion plots on the $g_{a \gamma \gamma}$ vs. axion--mass
plane.  The curve to the left are from Ref.~\protect\cite{la}. The letters indicate the following helium gas pressures in the conversion region of a 
strong magnetic field:  (a) vacuum, (b) 55~Torr, and (c) 100 Torr.}
\label{fig4}
\end{figure}

In Figure \ref{fig4} we show the area of the axion mass -- coupling constant
plane excluded by this result along with results of earlier work.

While this bound is interesting because it is a laboratory constraint,
it does not challenge the bound placed by Raffelt \cite{raff}, $g_{a
\gamma \gamma} \leq 10^{-10}$~GeV$^{-1}$ based on the helium burning
rate in low mass stars. A coupling constant $g_{a \gamma \gamma} \simeq
10^{-9}$~GeV$^{-1}$ would imply axion emission rates 100 times higher
than the stellar bound, and a significantly different concept of
stellar evolution.

This experiment can be considerably improved by using a large number
of smaller p--type germanium detectors with known orientations of the
(010) axes, with energy thresholds below 2~keV, and energy
resolutions corresponding to FWHM $\approx$ 0.5~keV. This is being
proposed at this time. Another collaboration could also operate the
COSME experiment in the new University of Zaragoza underground
laboratory in the Canfranc tunnel at 42$^{\rm o}$ 48'~N and 0$^{\rm o}$
31'~W.  The COSME detector is a 0.25~kg crystal having an energy
threshold of $\sim$1.8 keV and a resolution of 0.5~keV FWHM at
10~keV.   A positive result in this northern hemisphere experiment over
a wider energy range should have a very different temporal pattern from
that of the Sierra Grande experiment, but should yield the same value
of $\lambda$.

One of the USC/PNNL twin detectors is currently operating in the Baksan
Neutrino Observatory in Russia at 660~mwe, and if moved to a location
with greater overburden could also be used to acquire meaningful data
on solar axion rates.

The analysis presented here allows one to legitimately combine the
results of a number of experiments. Accordingly, the results from a
large number of experiments located throughout the world can be
combined to yield results equivalent to a single large experiment.

\end{document}